\documentclass[tradabstract]{aa}  

\usepackage{graphicx}
\usepackage{txfonts}
\usepackage{lipsum}
\usepackage{subcaption}         
                                
\usepackage{lscape}             
                               
\usepackage{placeins}            
                                
\usepackage{color}
\usepackage[dvipsnames]{xcolor}

\def\Gaia{{\it Gaia}}
\newcounter{Rco}
\newcommand{\Ionst}[1]{\setcounter{Rco}{#1}\Roman{Rco}}
\newcommand{\Ion}[2]{\mbox{#1\,{\scriptsize\Ionst{#2}}}}
\newcommand{\Ionw}[3]{\mbox{#1\,{\scriptsize\Ionst{#2}}~$\lambda\,#3$\,\AA}}

\newcommand{\Ionww}[3]{\mbox{#1\,{\scriptsize\Ionst{#2}}~$\lambda\lambda\,#3$\,\AA}}

\newcommand{\loggw}[1]{\mbox{$\log g\hspace{-0.5mm} =\hspace{-0.5mm} #1$}}

\newcommand{\Teff}{\mbox{$T_\mathrm{eff}$}\xspace}
\newcommand{\Teffw}[1]{\mbox{$\Teff\hspace{-0.5mm} =\hspace{-0.5mm} #1 \,\mathrm{kK}$}}
\newcommand{\ebv}{$E_\mathrm{B-V}$\xspace}

\newcommand{\Lsol}{$\mathrm{L}_\odot$\xspace}

\begin{document}

   \title{Ancient `ghost' planetary nebulae discovered with amateur telescopes}

   \subtitle{}

   \author{J.A. Manuel\inst{1}
        \and D. Jones\inst{2,3}
        \and M. Santander-Garc\'ia\inst{4}
        \and N. Reindl\inst{5}
        }

   \institute{Asociaci\'on Astron\'omica AstroHenares, Manuel Aza\~na, s/n, Coslada, Spain, \email{c3c273@gmail.com}
      \and
   Instituto de Astrof\'isica de Canarias, E-38205 La Laguna, Spain, 
   \and
    Departamento de Astrof\'isica, Universidad de La Laguna, E-38206 La Laguna, Spain
    \and
    Observatorio Astron\'omico Nacional (OAN-IGN), Alfonso XII, 3, 28014, Madrid, Spain
    \and
    Zentrum f\"ur Astronomie der Universit\"at Heidelberg, Landessternwarte, K\"onigstuhl 12, D-69117 Heidelberg, Germany
}

   \date{Received February 17, 2026}

  \abstract
  {As planetary nebulae evolve, they fade and dissipate into the surrounding interstellar medium, which makes them harder to detect. Modern, advanced amateur equipment can help to uncover this hidden population of ancient `ghost' planetary nebulae. Via careful processing of long-integration, narrow-band imagery with modest aperture telescopes at a dark-sky site, we detected three new candidate planetary nebulae (JAM~2, JAM~3, and JAM~4).  Each measures several arcminutes across with [O~\textsc{iii}] surface brightnesses of order 30 mag~arcsec$^{-2}$.  For each nebula, we identify a candidate central star, the parallaxes of which lead to nebular age estimates in the range 50--100 thousand years.  The candidate central star of JAM~2 also shows indications of photometric variability, potentially due to spots on the stellar surface.}


   \maketitle
    \nolinenumbers

\section{Introduction}

There have been many important amateur contributions to the field of planetary nebulae (PNe), particularly in the identification and confirmation of new candidates \citep[e.g.][]{2010PASA...27..156J,2012RMxAA..48..223A,2022A&A...666A.152L}. Additional compilations of confirmed and candidate PNe can be found in the Hong Kong/AAO/Strasbourg H$\alpha$ Planetary Nebula  (HASH PN) database and on the planetarynebulae.net website.\footnote{HASH \citep{2016JPhCS.728c2008P}: \url{http://hashpn.space/}; planetarynebulae.net: \url{https://planetarynebulae.net/EN/}.}These works highlight that the boundary between professional and advanced amateur astronomy has become increasingly blurred, particularly for low-surface-brightness phenomena. The combination of long integration times, narrow-band imaging, and stable, well-characterised instrumentation now allows small-aperture telescopes to probe surface brightness regimes that are largely inaccessible to wide-area professional surveys optimised for sky coverage rather than depth.  In this context, evolved (i.e. `ghost') PNe represent ideal targets for amateur telescopes \citep[see e.g.][]{2025arXiv250715834O}. The large angular sizes, extremely low surface brightnesses, and often fragmented or filamentary morphologies of these PNe make them difficult to detect in traditional surveys, yet they are well suited to deep, targeted observations by small telescopes operating under dark skies. In addition, the frequent dominance of [O~\textsc{iii}] emission and the weakness or absence of H$\alpha$ in these systems can hinder their identification when relying solely on existing survey data with limited sensitivity or filter combinations.

JAM~1 belongs to this emerging population of ancient PNe, with other recent examples presented in \citet{2022ApJ...935L..35F, 2025NatAs...9..380B,2025arXiv250715834O}. The discovery of JAM~1, first presented in \citet{2025RNAAS...9..251M}, illustrates how amateur-class facilities, when operated with professional-level methodologies, can play a decisive role not only in the detection but also the physical interpretation of such systems. By combining deep narrow-band imaging, astrometric information from \Gaia, and an analysis of the properties of the proposed central star, meaningful constraints can be placed on the evolutionary state of these faint remnants. This paper presents three further candidate ghost PNe (Table \ref{tab:jam}) identified using the same observational strategy and selection criteria. These objects share common characteristics, including extremely low surface brightness, large angular extents, and morphologies suggestive of advanced evolutionary stages. Together, they form a small but coherent sample that allows us to explore the observational properties of ghost PNe as a population, rather than as isolated discoveries.

This work therefore aims not only to report individual detections, but also to demonstrate that systematic searches for evolved PNe conducted with advanced amateur facilities can significantly complement professional surveys and contribute to a more complete census of the Galactic PN population, particularly at its faint and most elusive end.

\begin{figure*}
    \centering
    \includegraphics[width=0.7\textwidth]{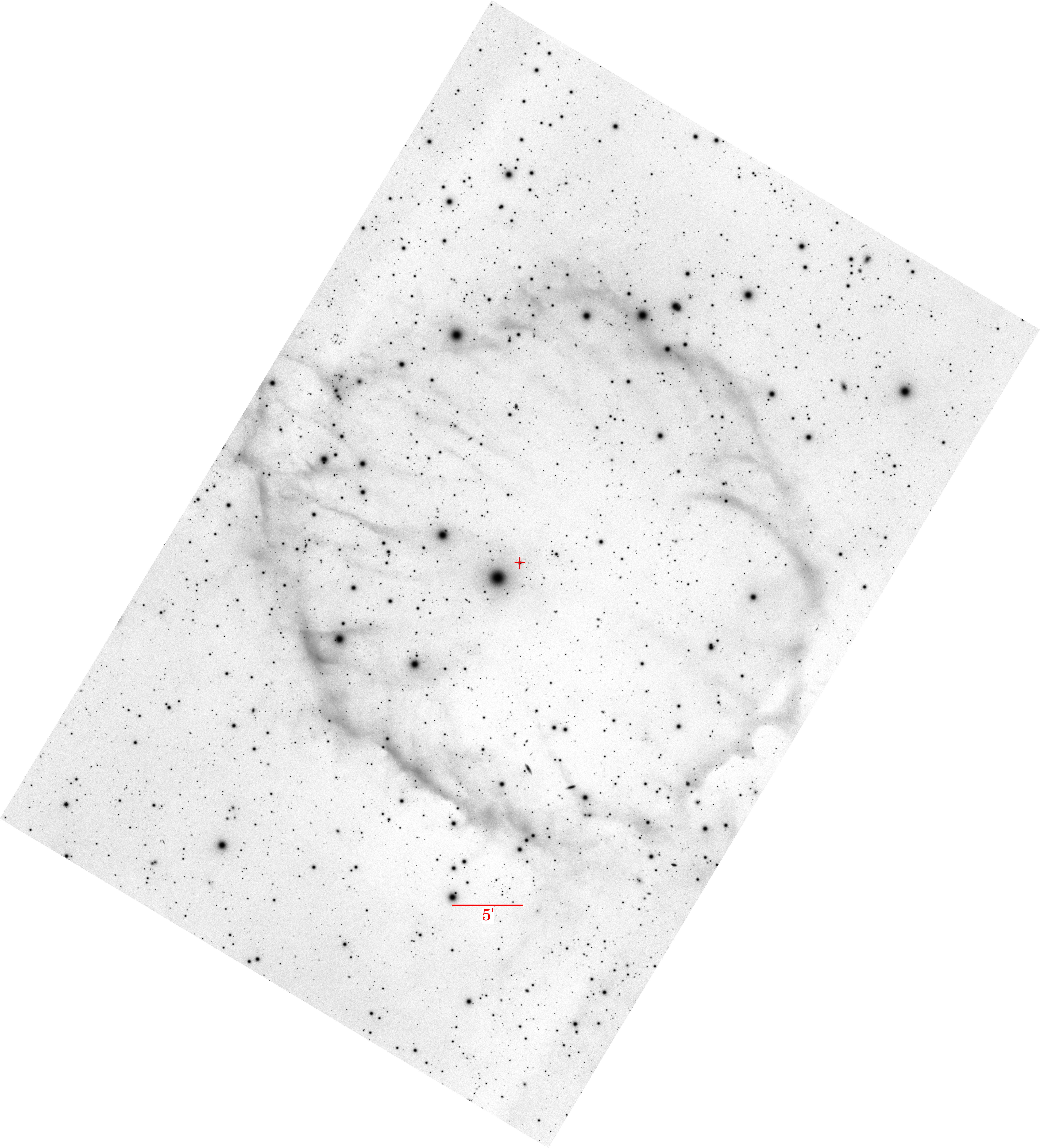}
    \caption{JAM~2 in the light of [O~\textsc{iii}]. The position of the candidate central star is marked with a red cross. North is up, and east to the left.}
    \label{fig:jam2_oiii}
\end{figure*}

\begin{table*}
\caption{New candidate ghost PNe.}
\centering
\begin{tabular}{l l l l c c l}
\hline
Name & Approximate geometric & Diameter & $t_{\mathrm{[O\,III]}}$ (h) & $t_{\mathrm{H\alpha}}$ (h) & Observing dates \\  & centre (J2000) &  &   &  &  \\
\hline 
JAM~1\tablefootmark{$\ast$} & 18:56:25 30:42:06 & 14.4\arcmin{} & -- & -- & --  \\
JAM~2 & 00:41:31 20:09:15 & 41\arcmin{}$\times$37\arcmin{}   &  131.0 & 47.8 & 2025 Aug 20--Dec 13 \\
JAM~3 & 19:19:37 73:39:52 & 7.1\arcmin{}$\times$5.9\arcmin{} &   49.3 & 31.0 & 2025 Oct 01--Nov 08 \\
JAM~4 & 07:58:20 66:46:07 & 9.6\arcmin{}$\times$8.3\arcmin{} &   48.5 & 19.5 & 2025 Nov 16--Dec 29 \\
\hline
\end{tabular}
\tablefoot{
\tablefoottext{$\ast$}{Previously reported in \cite{2025RNAAS...9..251M}.}
Total exposure times are given for each filter; all individual subframes had exposure times of 600\,s.}
\label{tab:jam}
\end{table*}

\begin{table*}
\caption{Candidate central stars.}
\centering
\begin{tabular}{l l c c c c} 
\hline
Name & Candidate central star & Geometric distance\tablefootmark{$a$} & \multicolumn{2}{c}{Proper motion\tablefootmark{$b$}} & Position angle\tablefootmark{$c$}\\
&&& $\mu_\alpha \cos\delta$ (mas/yr) & $\mu_\delta$ (mas/yr) &  (degrees)\\
\hline 
JAM~2 & Gaia DR3 2795519553550625536 & 397$\pm$6 pc & 25.1$\pm$0.1 & $-$11.53$\pm$0.07 & 114.7$\pm$0.1 \\
JAM~3 & Gaia DR3 2265067757936901760 & 2.1$^{+0.7}_{-0.5}$ kpc & $-$1.9$\pm$0.2 & $-$0.7$\pm$0.3 & 249$\pm$7\\
JAM~4 & Gaia DR3 1095335102795586944 & 1.05$^{+0.11}_{-0.10}$ kpc & 4.7$\pm$0.2 & 2.5$\pm$0.3 & 62$\pm$2\\
\hline
\end{tabular}
\tablefoot{
\tablefoottext{$a$}{\citet{2021AJ....161..147B}}
\tablefoottext{$b$}{Corrected for Galactic rotation following \citet{2023A&A...680A..99M}}
\tablefoottext{$c$}{The position angle of the proper motion vector following the correction for Galactic rotation}
}
\label{tab:jam_cs}
\end{table*}

\section{Survey strategy and observing setup}

The candidate evolved PNe presented here were identified through a systematic visual inspection of the publicly available narrow-band images from the astrophotographic survey created and maintained by Stefan Ziegenbalg\footnote{Public survey website: \url{https://www.simg.de}.}, aimed at detecting regions of excess [O~\textsc{iii}] emission. Objects exhibiting coherent, diffuse [O~\textsc{iii}] structures at very low surface brightness were flagged as candidates and subsequently followed up with deeper imaging.

Observations were carried out using a dual-telescope amateur facility located at a dark-sky site in Vald\'in (Ourense) in northern Spain. The instrumental setup consists of two identical apochromatic refractors operating in parallel, allowing simultaneous data acquisition and effectively doubling the total integration time for extended sources. Each telescope has an aperture of 150 mm with a focal ratio of f/7.1, providing a focal length of approximately 1073 mm. The two telescopes are equipped with identical monochrome CMOS cameras, each with a pixel scale of approximately $0.72\arcsec{}~\mathrm{pix}^{-1}$. This configuration is well matched to the large angular extent expected for highly evolved PNe.

\begin{figure*}
    \centering
    \includegraphics[width=\textwidth]{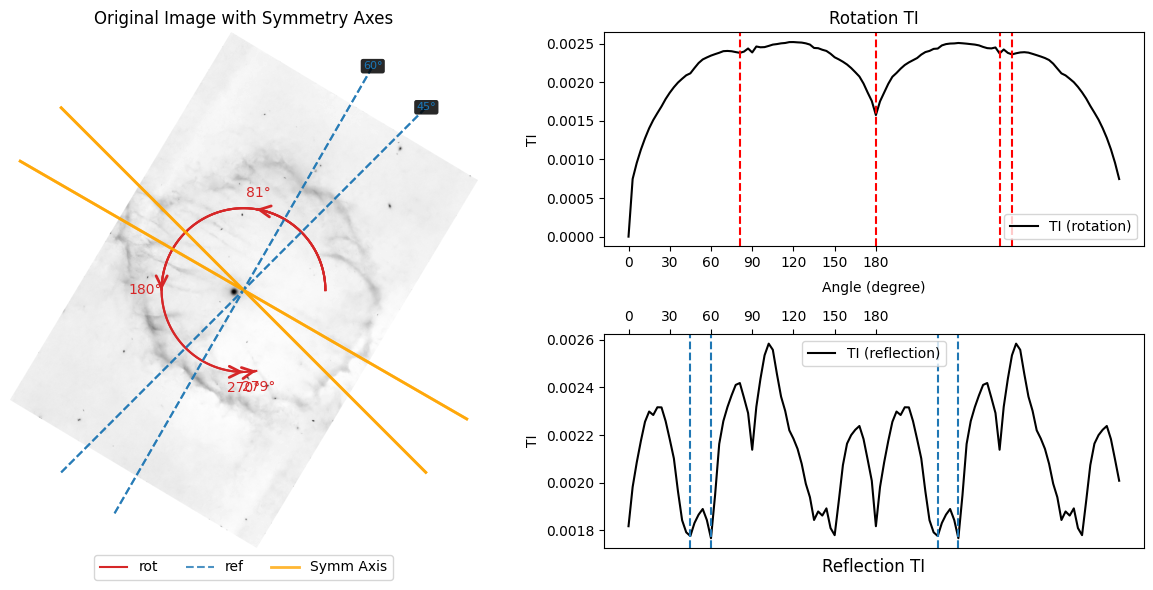}
    \caption{Starless [O~\textsc{iii}] image of JAM~2 and the resulting symmetry diagnostics.}
    \label{fig:sim_jam2}
\end{figure*}

Narrow-band imaging was performed using filters of Baader Planetarium CMOS optimised\footnote{https://www.baader-planetarium.com} centred on the [O~\textsc{iii}] $\lambda\,5007$ \AA{}  and H$\alpha$ emission lines, with bandwidths of  $40$ and $35$ \AA{}, respectively.  Here it is important to note that the H$\alpha$ filter also includes lines of [N~\textsc{ii}] in the wings of its transmission function -- lines that can be brighter than H$\alpha$ in evolved PNe.  However, in all cases, the [O~\textsc{iii}] emission was found to be significantly brighter and, thus, our analyses focus on these deeper images.

Additional broadband Sloan $g^{\prime}$ , $r^{\prime}$ and $i^{\prime}$ filters from the same manufacturer were used in selected cases to assess the presence of continuum emission and, to aid in the identification of potential central star candidates. The narrow-band observations were optimised for the detection of extremely low-surface-brightness emission, with individual exposure times ranging from several hundred to one thousand seconds per frame.

All observations were conducted under photometric or near-photometric conditions, with typical seeing values ranging from $1.2^{\prime\prime}$ to $2.3^{\prime\prime}$. Total integration times per target range from tens of hours to over one hundred hours when combining data from the two telescopes, placing these observations well beyond the depth of most existing wide-area surveys.

Initial image processing followed standard reduction procedures, including bias subtraction, dark-current correction, and flat-field normalisation. These steps were performed using a custom reduction pipeline developed in Python and based on the Astropy ecosystem, ensuring full control over calibration parameters and reproducibility of the reduction process. The resulting calibrated frames were then used for subsequent alignment, integration, and analysis.

Both optical trains were carefully characterised and maintained to ensure stable image quality and consistent photometric response throughout the observing campaigns. Dithering was applied between individual exposures to mitigate detector artefacts and improve background uniformity in the final combined images. The observing setup and reduction strategy are specifically optimised for the detection of extended, ultra-low-surface-brightness nebular emission and have proven particularly effective for identifying evolved or ghost PN candidates that are difficult or impossible to detect in shallower survey data.

\section{JAM~2}
JAM 2  is a very large, extremely low-surface-brightness nebula detected in deep [O~\textsc{iii}] imaging, located in the constellation of Pisces. The total integration time of the image shown in Fig.~\ref{fig:jam2_oiii}  was 131 h in [O~\textsc{iii}] obtained from 600s subframes \footnote{ At the time of submission of this manuscript, JAM~2 was not listed as a public object in the HASH PN database, nor did it appear on the planetarynebulae.net website. After email correspondence with HASH's database administrator, we were informed that an entry matching the position and angular size of JAM~2 had previously been inserted into database by Dr. D.J.~Frew, but had not been made public. We therefore report JAM~2 here as an independent identification, while acknowledging that the object had already been internally recorded in HASH.}.

The nebula exhibits an elliptical shell morphology with pronounced brightness asymmetries and filamentary substructure, consistent with a highly evolved PN interacting with the interstellar medium \citep[ISM;][]{2007MNRAS.382.1233W}. While the PN interpretation is favoured, confirmation will require kinematic or spectroscopic follow-up.

\subsection{Morphology}
The [O~\textsc{iii}] image reveals an elliptical shell with approximate angular dimensions of 41\arcmin{}$\times$37\arcmin{}. The shell is incomplete and highly non-uniform, with the emission concentrated into patchy, filamentary arcs rather than a continuous limb-brightened rim. This fragmented appearance is characteristic of advanced evolutionary stages, where hydrodynamic instabilities and ISM interaction affect the nebular morphology.
The brightest regions of the nebula exhibit a surface brightness of approximately $30\,\mathrm{mag\,arcsec^{-2}}$ in the [O~\textsc{iii}] band.

The brightest emission is confined to the eastern side of the shell, while the opposite side is fainter and more diffuse. The interior region is largely devoid of detectable emission, except for some wispy, filamentary patches, reinforcing the interpretation of a thin, expanded shell rather than a filled H~\textsc{ii} region or diffuse Galactic emission. No comparable large-scale structure is visible in adjacent fields at similar depths, arguing against a chance superposition of unrelated ISM filaments.

\begin{figure*}[t]
    \centering
    \includegraphics[width=\textwidth]{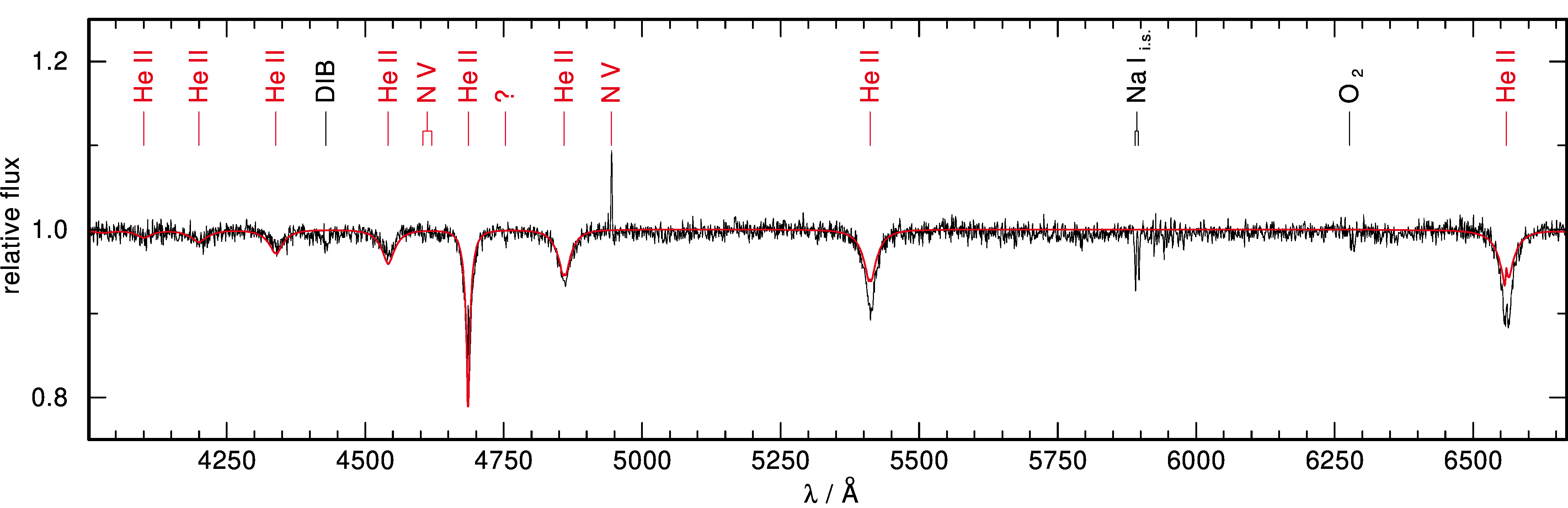}
    \caption{GTC-OSIRIS spectrum of the candidate central star of JAM~2, PG~0038+199. Over-plotted is a H+He TMAP model with \Teffw{125}, \loggw{7.0}, and H/He$=0.02$ (by mass).}
    \label{fig:JAM2_spectrum}
\end{figure*}

\begin{figure}
    \centering
    \includegraphics[width=1.0\linewidth]{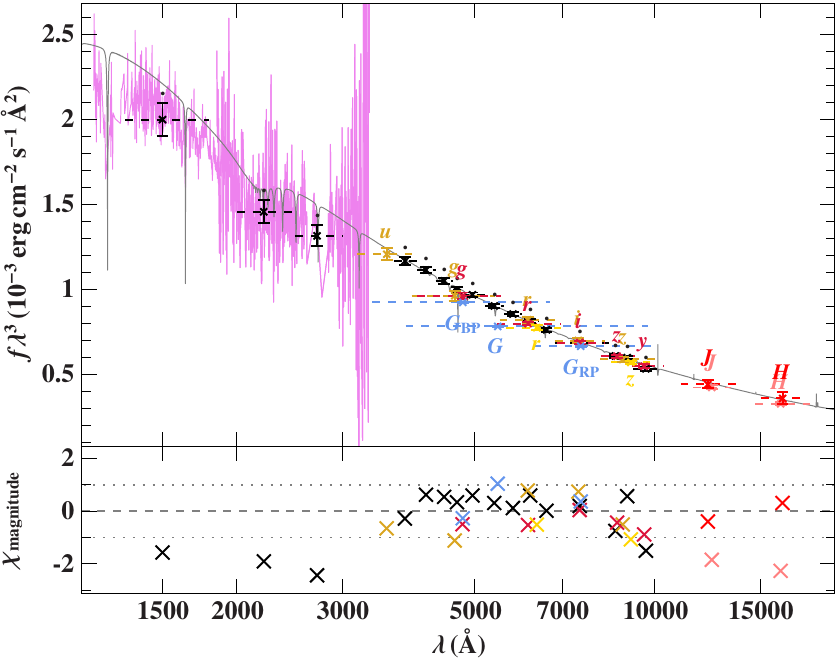}
    \caption{Fit to the SED for PG\,0038+199. Top panel: Filter-averaged fluxes converted from observed magnitudes are shown in different colours. The pink line represents the IUE spectra, and the grey lines correspond to the model flux that best fits the observation. The model flux is degraded to a spectral resolution of $6\,\AA$. To reduce the steep SED slope, the flux is multiplied by the wavelength cubed. Bottom panel: Difference between synthetic and observed magnitudes.}
    \label{fig:JAM2_sed}
\end{figure}

To quantify the symmetry properties of the [O~\textsc{iii}] emission in our new nebular candidates, we adopted the transformation information (TI) formalism introduced by \citet{2026arXiv260107913S}. In this approach, the similarity between an image, $I(\mathbf{x})$, and a geometrically transformed version, $T_\theta I(\mathbf{x})$ (either a rotation by angle $\theta$ or a reflection about an axis at an angle $\theta$) is measured through the TI statistic computed over the overlap domain between the original and transformed images. Local minima in the resulting TI$(\theta)$ curves identify candidate rotational symmetries (rotation TI) and mirror-symmetry axes (reflection TI), allowing a robust, metric-based characterisation of symmetry even in low-surface-brightness regimes.

To enhance the visibility of the extremely low-surface-brightness nebular emission and to prevent field stars from dominating the morphological and symmetry diagnostics, we performed the analysis on star-subtracted (`starless') versions of the [O~\textsc{iii}] images. The starless frames were produced with \textsc{StarXTerminator} \citep{StarXTerminator} software. This step reduces the risk of spurious symmetry signatures driven by the spatial distribution of field stars rather than by the nebular emission itself.

The starless image of JAM~2 (Fig.~\ref{fig:sim_jam2}, left) reveals a large, evolved shell with a non-uniform rim brightness, yet the TI diagnostics indicate a coherent global symmetry component. The rotation TI curve exhibits its most pronounced non-trivial minimum at $\theta \simeq 180^{\circ}$ (excluding the trivial $\theta\rightarrow 0^{\circ}$ self-match), implying that the [O~\textsc{iii}] surface-brightness distribution is significantly more self-similar under a half-turn rotation than under neighbouring angles. We could interpret this as evidence of an axisymmetry in projection, rather than an exact two-fold symmetry. Additional, shallower rotation minima are present (e.g.\ near $\theta \simeq 81^{\circ}$ and $\theta \simeq 270$--$280^{\circ}$), and likely reflect weaker, localised self-similarity associated with segmented rim structure and patchy low-surface-brightness emission.

The reflection TI curve identifies a small set of candidate mirror-symmetry solutions, with prominent minima clustered around $\theta_{\rm ref}\approx 45^{\circ}$--$60^{\circ}$ \citep[note that, as in][the angle $\theta_{\rm ref}$ is defined as the anti-clockwise angle from the $x$-axis of the image, which in our case corresponds to the position angle plus 90$^\circ$]{2026arXiv260107913S}, corresponding to projected mirror axes on the sky within a narrow angular range. This clustering suggests a preferred morphological orientation (a principal symmetry plane in projection), while the departures from perfect reflection symmetry are dominated by brightness variations along the shell rim and diffuse extensions at very low surface brightness. Given that TI-based symmetry signatures can be affected by the adopted centre and residual large-scale background structure in such faint regimes, we conservatively report the $\sim 180^{\circ}$ rotation minimum and the clustered reflection-axis solutions as the most robust symmetry features of JAM~2 in the present data.

\subsection{Central star}
JAM~2 seems to be spatially associated with the well-studied DO1 white dwarf (WD), PG 0038+199, very close to the geometric centre of the nebula ($\alpha = 00^{\rm h}41^{\rm m}35.338^{\rm s}$, $\delta = +20^{\circ}09^{\prime}16.907^{\prime\prime}$, J2000).  \Gaia\ Data Release 3 (DR3) astrometry gives a parallax of $2.50 \pm 0.04\,\mathrm{mas}$ \citep{2023A&A...674A...1G}, corresponding to a distance of $397 \pm 6\,\mathrm{pc}$ \citep{2021AJ....161..147B}. 

The star has $G = 14.491 \pm 0.003\,\mathrm{mag}$, $BP-RP = -0.53 \pm 0.05\,\mathrm{mag}$, and an absolute magnitude $M_G \approx 6.48 \pm 0.04\,\mathrm{mag}$, fully consistent with a hot, luminous WD.  Its proper-motion vector is approximately aligned with the minor axis of the nebula, and the star is moving towards the brighter, compressed side of the shell, as clearly indicated in Fig.~\ref{fig:jam2_oiii}. This configuration is a well-known observational signature of PN–ISM interaction, where ram pressure enhances emission on the leading edge and erodes the trailing side \citep{2007MNRAS.382.1233W}.

\begin{figure}
    \centering
    \includegraphics[width=1.0\linewidth]{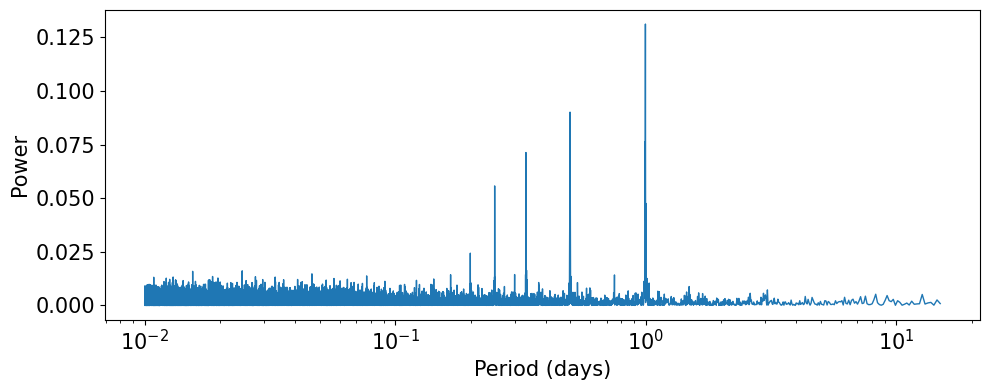}
    \caption{Lomb-Scargle periodogram of the zg light curve of PG~0038+199. The maximum power is at 0.994 d.}
    \label{fig:JAM2_periodogram}
\end{figure}

\begin{figure}
    \centering
    \includegraphics[width=1.0\linewidth]{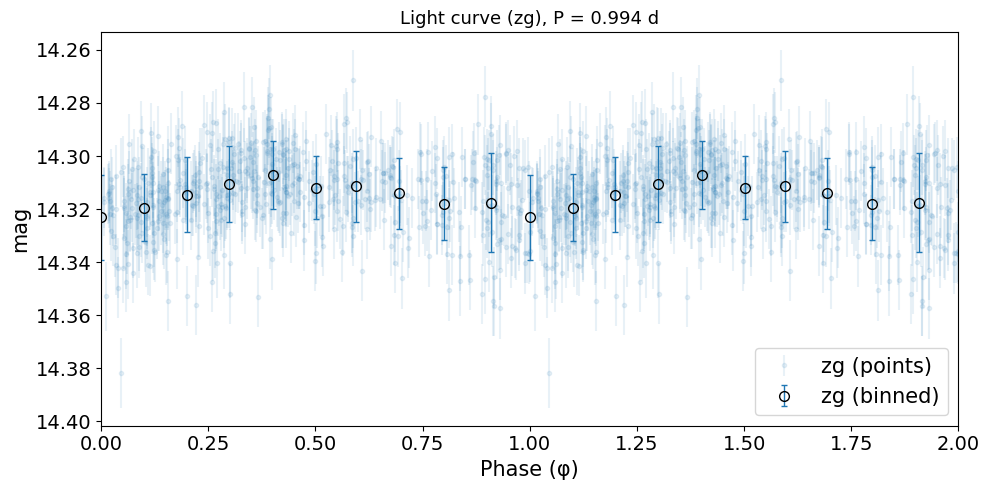}
    \caption{zg light curve of PG~0038+199 folded on the peak of the Lomb-Scargle periodogram (0.994 d).}
    \label{fig:WD_JAM2_lc}
\end{figure}

At the \Gaia\ DR3 distance of 400 pc, the angular semi-major axis corresponds to a physical radius of R $ \approx 2.4 \ $ pc, placing JAM~2 among the largest PN candidates known. Assuming a generous expansion velocity of $V_{\rm exp} \simeq 30\,\mathrm{km\,s^{-1}}$, the inferred kinematical age is $t \approx 80\,000 \,\mathrm{yr}$, consistent with a late-stage PN approaching full dispersal into the ISM.

PG~0038+199 is regarded as a canonical standard star for the hot, helium-rich DO1 spectral class \citep{1993PASP..105..761W}. The first non-local thermodynamics equilibrium (NLTE) spectral analyses of PG~0038+199 were presented by \cite{1996A&A...314..217D} and \cite{1997fbs..conf..303D}, who found \Teffw{115} and \loggw{7.5} based on optical TWIN and KECK spectroscopy, respectively. A sophisticated combined optical and ultraviolet spectral analysis of PG~0038+199 was presented by \cite{2017A&A...601A...8W}, who derived \Teffw{125\pm5} and \loggw{7.0\pm0.5}. In addition they found that PG\,0038+199 shows remaining H (2\%, by mass), has a super solar N abundance ($30\times$ solar), but subsolar abundances of C and O. 

Spectra were obtained on August 26, 2023, using the Optical System for Imaging and low-Intermediate-Resolution Integrated Spectroscopy (OSIRIS) mounted on the 10.4m Gran Telescopio Canarias (GTC). The R2000B grating and R2500R grating were used along with a 0.6" long slit to obtain single exposures of 450s (programme ID: GTC51-23B).
Besides the previously identified \ion{He}{ii} absorption and \Ionw{N}{5}{4945} emission, we see hints of absorption lines of the \Ionww{N}{5}{4604, 4620} doublet (see Fig.~\ref{fig:JAM2_spectrum}). \cite{2023MNRAS.519.2321J} show that around \Teff$\approx 120$\,kK, the \Ionww{N}{5}{4604, 4620} doublet becomes invisible as the lines turn from absorption to emission. In addition we detect an unidentified absorption line around 4753\,\AA{}  and a weak diffuse interstellar band at 4429\,\AA.

As already reported by \cite{1996A&A...314..217D}, the star exhibits a \Ion{He}{2} line problem, meaning that not all \Ion{He}{2} lines can be fit simultaneously. This also becomes obvious in Fig.~\ref{fig:JAM2_spectrum} where we have over-plotted a H and He model which we calculated using the T{\"u}bingen NLTE Model-Atmosphere Package (TMAP\footnote{\url{http://astro.uni-tuebingen.de/~TMAP}}; \citealt{tmap2012}) assuming the best-fit parameters (\Teffw{125\pm5}, \loggw{7.0\pm0.5}, H/He$=0.02$) from \cite{2017A&A...601A...8W}. The blue part of the spectrum is well reproduced, except the central emission of \Ionw{He}{2}{4686}. On the other hand, the observed \Ionww{He}{2}{5411, 6560} absorptions are too deep compared to the model.

We performed a fit to the observed spectral energy distribution (SED), employing the pure He TMAP model grid presented by \cite{Reindl+2023} and the $\chi ^2$ SED fitting routine described in \cite{Heber+2018} and \cite{Irrgang+2021}. 
We used photometry from \Gaia\/ (Early)DR3 \citep{Gaia+2020, Gaia+2021}, photometry from various surveys, as well as the IUE spectra of the star. In the fit we assumed the best fit parameters (\Teffw{125\pm5}, \loggw{7.0\pm0.5}) from \cite{2017A&A...601A...8W} and let only the angular diameter, $\Theta$, and the colour excess vary freely. Interstellar reddening was accounted for by using the reddening law of \cite{Fitzpatrick+2019} with $R_V=3.1$.

Our best fit is shown in Fig.~\ref{fig:JAM2_sed}. The UV flux is slightly overestimated by the model, because of the lack of metal opacities which would decrease the predicted UV flux.
We derive a reddening of \ebv$=0.033\pm0.002$\,mag, in agreement with the 2D dust map of \cite{Schlafly2011} who predicts \ebv$=0.033\pm0.002$\,mag. From the angular diameter and parallax, the radius of the stars can be calculated via \textit{R} = $\Theta$/2$\varpi$. By that we obtain $R_{\mathrm{Gaia}}=0.0285 \pm 0.0013$\,$R_\odot$, and using $L/L_\odot = (R/R_\odot)^2(T_\mathrm{eff}/T_{\mathrm{eff},\odot})^4$ a luminosity of $L= 180 \pm 40$\,\Lsol.

\begin{figure*}
    \centering
    \includegraphics[width=0.7\textwidth]{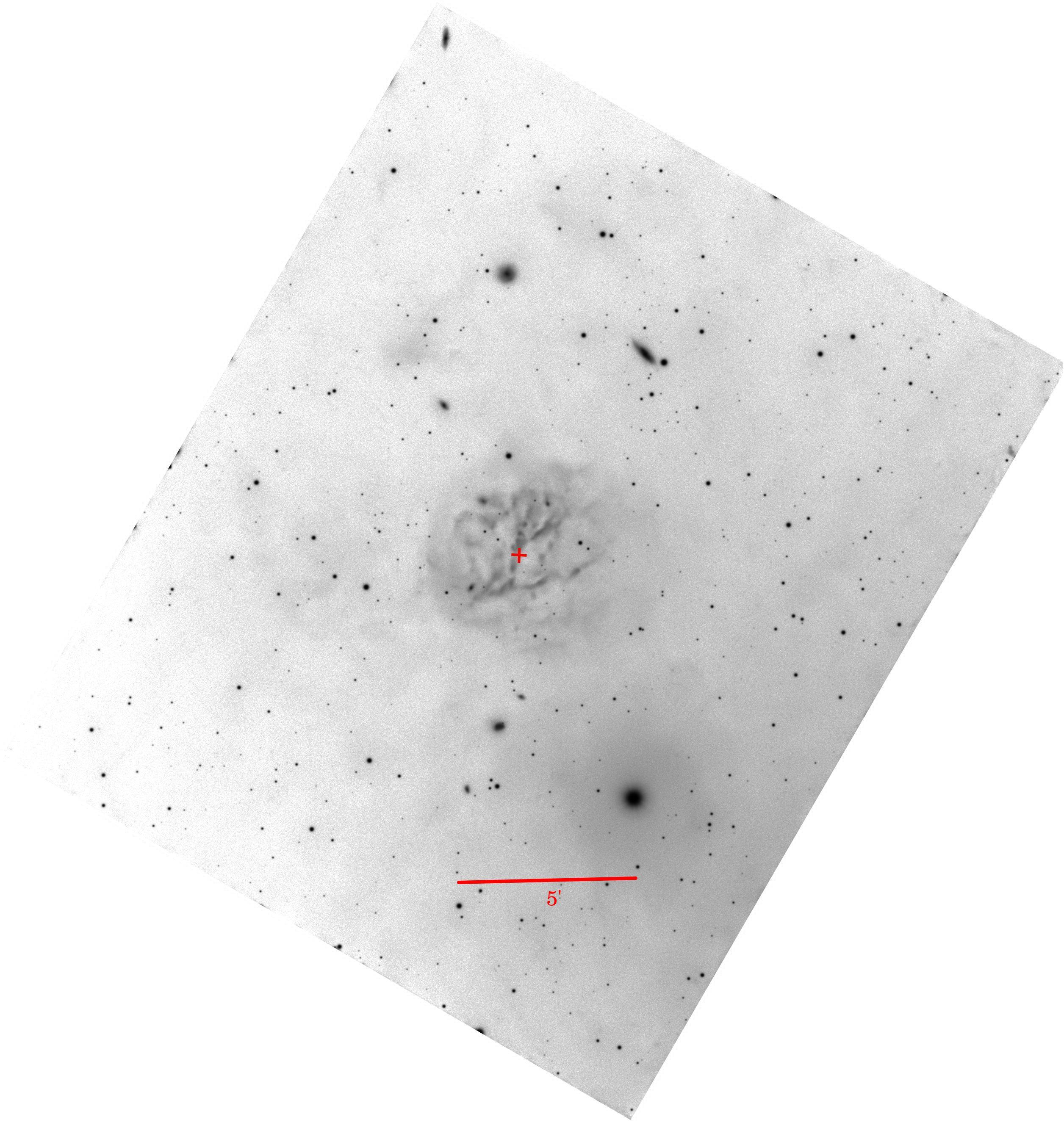}
    \caption{JAM~3 in the light of [O~\textsc{iii}]. The position of the candidate central star is marked with a red cross. North is up, and east to the left.}
    \label{fig:jam3_oiii}
\end{figure*}

\begin{figure*}
    \centering
    \includegraphics[width=\textwidth]{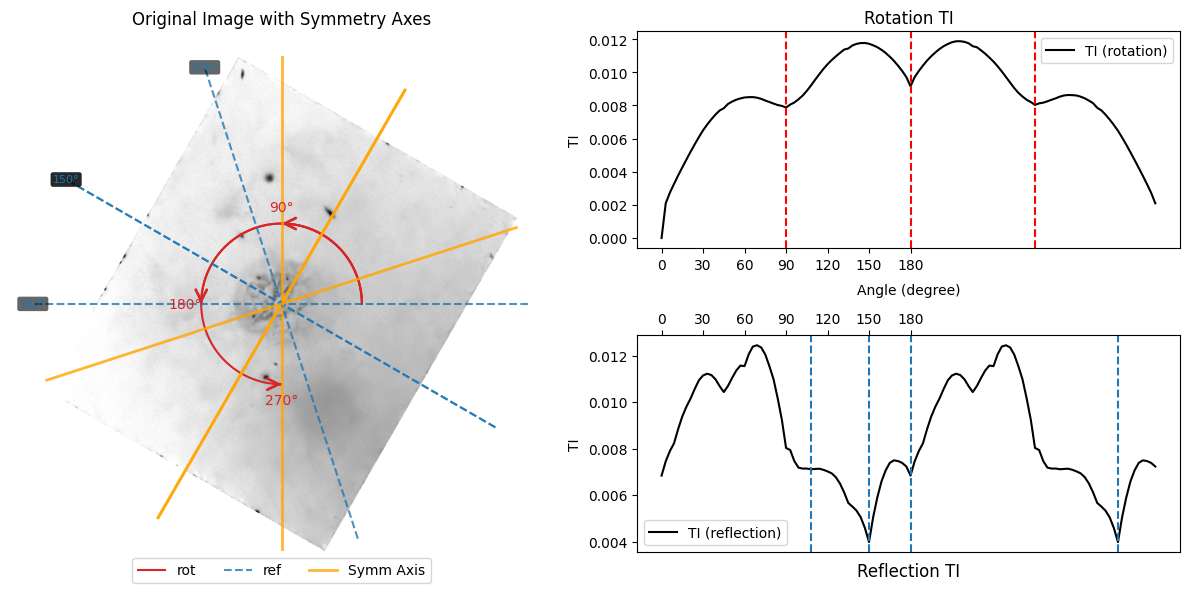}
    \caption{Starless [O~\textsc{iii}] image of JAM~3 and the resulting symmetry diagnostics. }
    \label{fig:sim_jam3}
\end{figure*}

Analysis of Zwicky Transient Facility \citep[ZTF;][]{ztf} light-curve data for the candidate central star reveals statistically significant photometric variability with a period of approximately $0.994 \pm 0.002 \,\mathrm{d}$ , as indicated by the periodogram and the phase-folded zg-band light curve shown in Figs.~\ref{fig:JAM2_periodogram} and ~\ref{fig:WD_JAM2_lc}, respectively. The ZTF zr-band light curve exhibits the same behaviour, with a consistent period within the uncertainties.  The period is suspiciously close to one day, potentially indicative of an alias of the sampling. However, the variability is not correlated with the airmass, seeing or lunar phase of the observations, providing some indication that the variability is intrinsic and not an artefact of reduction or photometry. A by-eye inspection of the folded light curve in Fig.~\ref{fig:WD_JAM2_lc} shows that the brightness varies relatively smoothly in a roughly sinusoidal manner, with an amplitude of a few tens of millimagnitudes. The derived ephemeris for the heliocentric Julian date (HJD) of minimum light is

\begin{equation}
\mathrm{HJD}_\mathrm{min}=2458271.02(27) + 0.994(2) E,
\end{equation}

where $E$ is an integer that represents the number of cycles since the time of reference minimum.

PG~0038+199 was also observed by TESS (TIC~434216729) in two sectors with the data showing no notable periodicity perhaps indicating that the ZTF variability is, in fact, simply an alias.  However, the CROWDSAP is quite low at 0.64 (indicating that some 36\% of the observed flux does not originate from PG~0038+199, diluting any potential variability) and the star is not particularly bright  ($\sim$1~magnitude fainter than the typically quoted primary magnitude limit of the mission).  

The origin of the ZTF variability (assuming it is real) is unclear but its apparently sinusoidal nature could be consistent with irradiation of an otherwise hidden cool companion.  However, no irradiated lines (originating from the surface of a cool companion) are visible in the spectra \citep[see e.g.][]{2020A&A...642A.108J}.  Furthermore, the SED shows no indication of an infrared excess (Fig.~\ref{fig:JAM2_sed}) making a main sequence companion unlikely.  To test what type of non-degenerate companion could potentially be hidden in the SED, we employed the PHOENIX model grid \citep{2013A&A...553A...6H} and adopted representative effective temperatures and radii for M dwarfs from  \cite{2020A&A...642A.115C}. We find that companions earlier than spectral type M6 would be detectable in the SED.

A WD companion is also unlikely as the period is too long to be associated with ellipsoidal modulation (where the true period would be double the observed peak), and this would imply a radius (for either component) inconsistent with the SED fitting.  The most likely cause of the variability is thus rotation and spots \citep[see e.g.][]{2020NatAs...4.1092M,2021A&A...647A.184R}.  Comparing the two ZTF light curves, there is no appreciable change in colour index (g-r) as a function of phase, a typical indication of spots, but the low amplitude and relatively low signal-to-noise of the variability makes this inconclusive. Ultimately, further confirmation via high-precision multi-band times-series photometry and/or spectroscopy is required in order to confirm the variability and its possible origin.

Besides PG~1034+001, PG~0038+199 is currently the only N-rich DO WD. These stars are thought to be the direct successors of the N-rich O(He) stars, which exhibit similar CNO abundance patterns \citep{2017A&A...601A...8W}. Currently, there are fourteen O(He) stars known \citep{2014A&A...566A.116R, 2014A&A...564A..53W, 2015A&A...583A.131W, 2022A&A...658A..66W, 2023MNRAS.519.2321J, 2025A&A...693A.167W}, nine of which belong to the N-rich subtype. Of these nine stars, three are central stars of PNe (K~1-27, LoTr~4, and Pa~5), and no PN has been reported around the other six. Depending on whether a (H-rich) PN is present or not, two different evolutionary pathways have been proposed. Naked (i.e. no PN) N-rich O(He) stars are thought to be the product of a double He-core WD merger \citep{2014A&A...566A.116R, Zhang+2012a}, and the direct successors of the He-core burning and N-rich He-sdO stars. For O(He) stars that are surrounded by a H-rich PN, the double He-core WD merger scenario is not possible, given the long ($>10^7$\,yrs) He-core burning phase. For these central stars it has been suggested that they might result from a merger of an asymptotic giant branch star within a common envelope, meaning that the ejected envelope is visible as H-rich PN, while the merger might have produced the H-deficiency of the central star \citep{2014A&A...566A.116R}.  
The evolutionary status of the two N-rich DO WDs would then depend on whether the ionised material around PG~0038+199 and PG~1034+001 can be associated with of a H-rich PN or not. We, also note that the evolutionary models of \citet{2009ApJ...704.1605A} indicate that PG~0038+199 is already 40--50 kyr along its cooling track, consistent with being inside an ancient PN \citep[the time between PN formation and the star reaching maximum effective temperature is uncertain, but can range from a few hundred to a few tens of thousands of years;][]{2016A&A...588A..25M}.

\section{JAM~3}
JAM~3 is a low-surface-brightness nebula located in the constellation of Draco, detected in deep narrow-band [O~\textsc{iii}] imaging. The [O~\textsc{iii}] frame shown in Fig.~\ref{fig:jam3_oiii} was obtained from 296 subframes of 600\,s each (total integration time 49.3\,h) and is oriented with north up and east to the left (plate scale $0.72\arcsec\,\mathrm{px}^{-1}$).

\begin{figure*}
    \centering
    \includegraphics[width=0.7\textwidth]{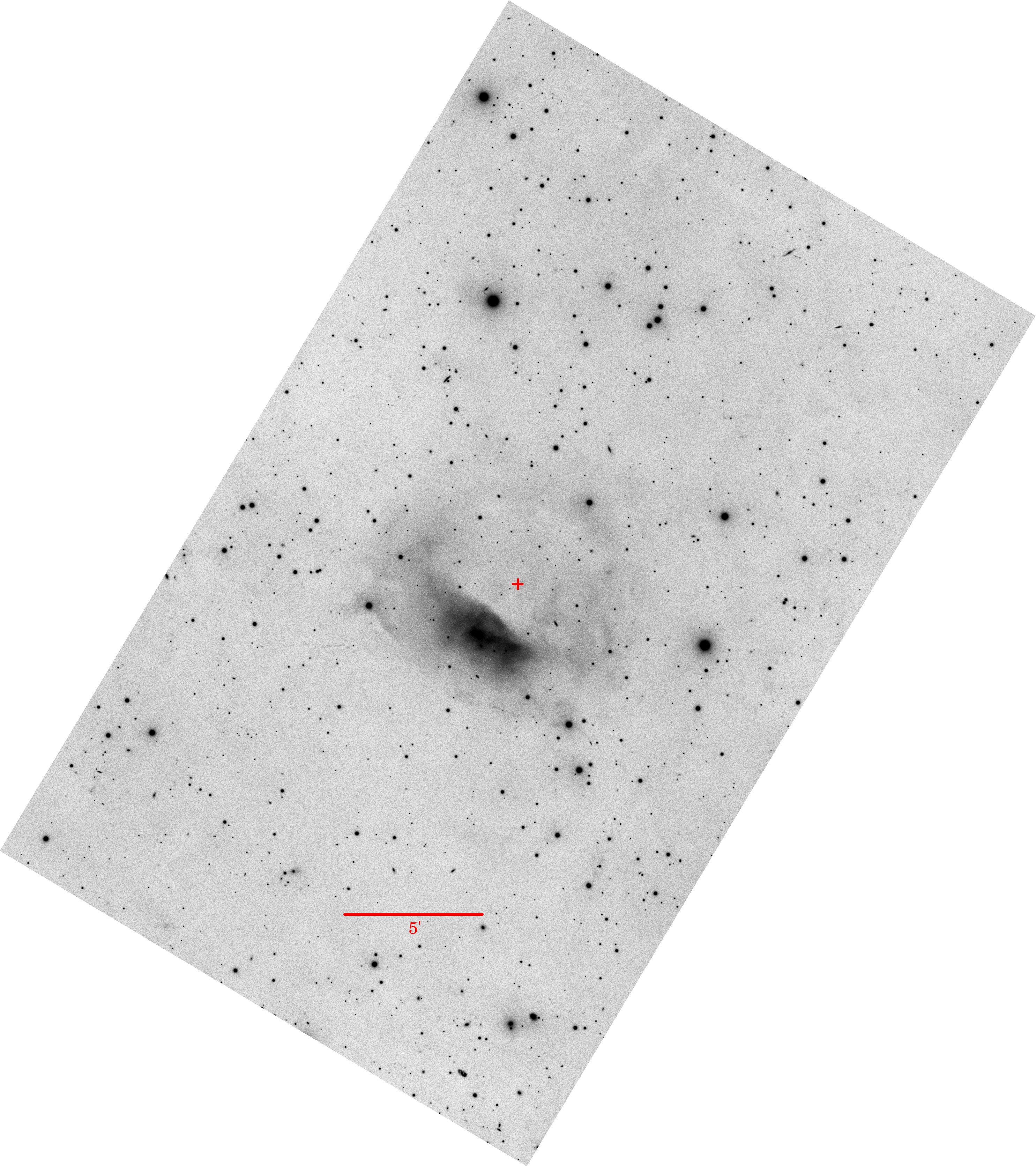}
    \caption{JAM~4 in the light of  [O~\textsc{iii}]. The position of the candidate central star is marked with a red cross. North is up,and east to the left.}
    \label{fig:jam4_oiii}
\end{figure*}

\begin{figure*}
    \centering
    \includegraphics[width=\textwidth]{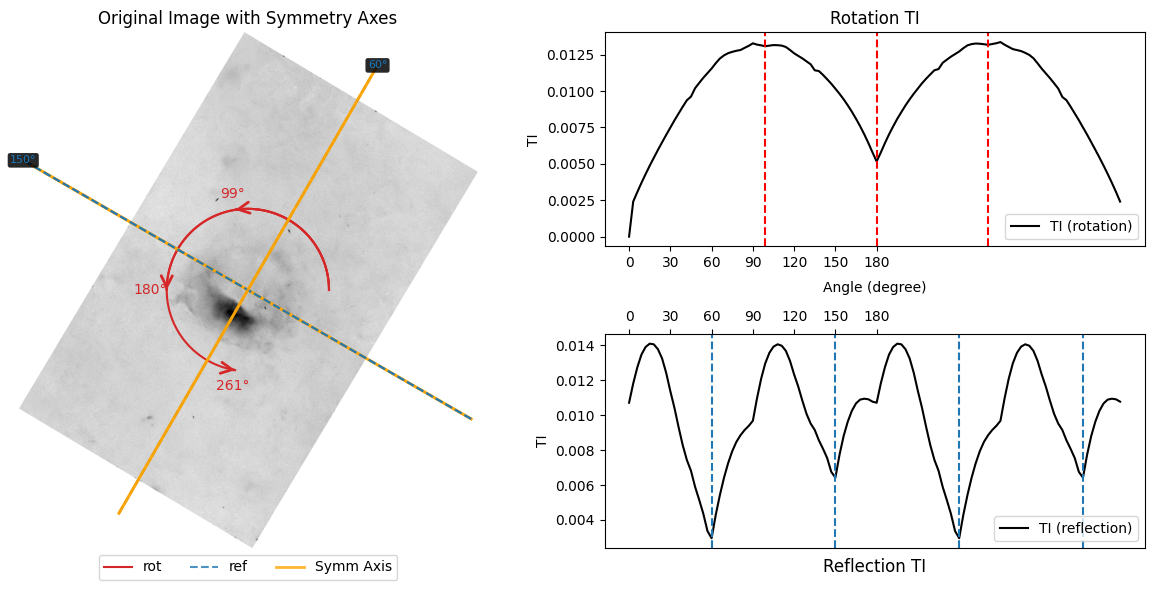}
    \caption{Starless [O~\textsc{iii}] image of JAM~4 and the resulting symmetry diagnostics.}
    \label{fig:sim_jam4}
\end{figure*}

\subsection{Morphology}
Morphologically, JAM~3 exhibits an overall elliptical appearance that is largely filled with prominent filamentary and arc-like substructure across the interior, rather than a clean, centrally evacuated limb-brightened ring. The highest-contrast features trace a network of internal filaments and partial rim segments, while a fainter, more diffuse outer envelope extends beyond the structured inner emission. 
From an isophotal segmentation of the extended [O\,\textsc{iii}] emission, adopting an isophote at 25\% of the image peak intensity (i.e. $I_{\rm iso}=0.25\,I_{\max}$), the outer envelope spans approximately $D_{\rm maj}\simeq592$ pix and $D_{\rm min}\simeq491$ pix, corresponding to $7.1' \times 5.9'$.

In the starless [O~\textsc{iii}] image of JAM~3, it appears as a low-contrast, diffuse structure close to the detection limit over much of the field, and its TI signatures are correspondingly dominated by large-scale morphology. In this regime, the TI diagnostics are expected to be more sensitive to low-frequency structure (and potentially to residual background gradients) than in well-defined shell sources. The rotation TI curve shows multiple non-trivial local minima, including candidates near $\theta \simeq 90^{\circ}$, $\theta \simeq 180^{\circ}$, and $\theta \simeq 270^{\circ}$, indicating that the image exhibits comparable degrees of self-similarity under several rotations. The presence of these minima of similar character argues against a uniquely dominant rotational symmetry at the current signal-to-noise level.

The reflection TI curve likewise identifies several candidate mirror-symmetry solutions (with minima distributed over multiple angles), rather than a single sharply preferred axis. We therefore conservatively interpret the reflection results as evidence of weak or ambiguous mirror symmetry in JAM~3, compatible  either with intrinsically irregular or low-axisymmetry structure or with the limited contrast of the detected emission. Given the known sensitivity of TI-based symmetry signatures to the adopted centre and large-scale background residuals in very faint regimes, we treated the set of detected rotation and reflection minima as candidate symmetries, and refrained from assigning a unique principal symmetry axis for JAM~3 from the current image alone; see Fig.~\ref{fig:sim_jam3}.

\subsection{Central star}
Near the geometric centre of the nebula we identify a star classified as a WD candidate \citep{2021MNRAS.508.3877G,2022A&A...662A..40C}, Gaia DR3 2265067757936901760 ($\alpha = 19^{\rm h}19^{\rm m}42.35^{\rm s}$, $\delta = +73^{\circ}40^{\prime}01.55^{\prime\prime}$; J2000). The \Gaia\ DR3 parallax, $\varpi = 0.5448 \pm 0.1149\,\mathrm{mas}$, implies a distance estimate of $d = 2.1^{+0.7}_{-0.5}\,\mathrm{kpc}$. The source has $G = 18.432 \pm 0.003\,\mathrm{mag}$ and $BP-RP = -0.297 \pm 0.044\,\mathrm{mag}$, consistent with a hot blue object, and yields an absolute magnitude of $M_G = 7.11 \pm 0.46\,\mathrm{mag}$. Using the measured major angular diameter of the nebula ($D_{\rm maj}=7.1'$) and the inferred distance, the corresponding physical major diameter is $D_{\rm maj,phys} = 4.3^{+1.4}_{-1.0}\,\mathrm{pc}$ . Adopting a representative expansion velocity of $v_{\rm exp}=30\,\mathrm{km\,s^{-1}}$ for an evolved PN, we obtain a kinematic age of $t_{\rm kin} = 71^{+24}_{-17} \, \mathrm{kyr}$.

We searched for photometric variability of the candidate central star using publicly available ZTF time-series photometry in the $g$ and $r$ bands. We analysed the light curves after standard quality filtering and performed a period search using a Lomb--Scargle periodogram over periods spanning $\sim$0.01--15\,d. No statistically significant periodic modulation was detected in either band, and the phase-folded light curves do not show the coherent sinusoidal (reflection or ellipsoidal) signatures typically associated with short-period post-common-envelope binaries. These results therefore do not support a close-binary interpretation based on ZTF photometry alone, although low-amplitude variability, unfavourable inclination, or periods poorly sampled by the ZTF cadence cannot be excluded.

\section{JAM~4}
JAM~4 is a  low-surface-brightness nebula in the constellation of Camelopardalis, detected in deep narrow-band [O~\textsc{iii}] imaging. The [O~\textsc{iii}] data shown in Fig.~\ref{fig:jam4_oiii} were obtained from 291 subframes of 600\,s each (total integration time 48.5\,h); the image is oriented with north up and east to the left. During the preparation of this manuscript, the discovery of JAM~4 was independently reported by \citet{2026AN....34770078C} during a survey for dwarf companions to the spiral galaxy NGC~2403.  They dub the nebula TBG-N1.

\subsection{Morphology}
Morphologically, JAM~4 is shell-like and predominantly limb-brightened, with a markedly brighter sector along the rim, plausibly produced by compression and enhanced excitation where the expanding shell interacts with the ambient ISM (PN--ISM interaction). 

The starless image of JAM~4 is dominated by a compact, high-contrast central emission component superposed on much fainter large-scale structure. In this situation, the TI diagnostics primarily probe the symmetry content of the dominant central brightness distribution rather than that of any extremely low-surface-brightness outer emission. The rotation TI curve displays a pronounced non-trivial minimum at $\theta \simeq 180^{\circ}$ (excluding the trivial $\theta\rightarrow 0^{\circ}$ self-match), indicating that the [O~\textsc{iii}] surface-brightness pattern is significantly more self-similar under a half-turn rotation than under neighbouring angles. We interpret this as evidence of a dominant centro-symmetric (C$_2$-like) component in projection, while noting that the object is not strictly two-fold symmetric owing to the marked brightness asymmetry of the compact core. Additional, shallower rotation minima (e.g. near $\theta \simeq 99^{\circ}$ and $\theta \simeq 261^{\circ}$) likely trace weaker, localised self-similarity associated with anisotropic extensions around the core.

The reflection TI curve identifies well-defined candidate mirror-symmetry solutions, with prominent minima consistent with reflection axes near $\theta_{\rm ref}\approx 60^{\circ}$ and $\theta_{\rm ref}\approx 150^{\circ}$. This indicates a preferred morphological orientation (a principal symmetry plane in projection) for the dominant emission component, with departures from perfect mirror symmetry driven by one-sided brightness enhancements in the compact core and by low-contrast diffuse structure. Given that TI-based signatures can be sensitive to the adopted centre, masking, and residual large-scale background structure, we conservatively report the $\sim 180^{\circ}$ rotation minimum and the identified reflection-axis candidates as the most robust symmetry features of JAM~4 in the current data. See Fig.~\ref{fig:sim_jam4}.

\subsection{Central star}
Near the nebular centre we identify a blue \Gaia\ WD candidate, Gaia DR3 1095335102795586944, located at $\alpha = 07^{\rm h}58^{\rm m}20.029^{\rm s}$, $\delta = +66^{\circ}45^{\prime}58.68^{\prime\prime}$ (J2000), marked by a red cross in Fig.~\ref{fig:jam4_oiii}. The \Gaia\ parallax of $\varpi = 0.98\,\pm\,0.09\,\mathrm{mas}$ implies a distance of $d = 1.05^{+0.11}_{-0.10}\,\mathrm{kpc}$. The \Gaia\ proper motion vector once corrected for Galactic rotation (Table \ref{tab:jam_cs}) points approximately east-northeast, roughly perpendicular to the brightened edge or rim.  If this brightening were the result of interaction with the ISM, one would expect the direction of proper motion to be roughly to the south-east, thus calling this hypothesis into question. Adopting a representative expansion velocity of $v_{\rm exp}=30\,\mathrm{km\,s^{-1}}$, the implied kinematic age based on the nebular major-axis radius is $t_{\rm kin} = 48\pm5 \,\mathrm{kyr}$.
The source has $G = 17.442\,\pm\,0.003\,\mathrm{mag}$ and $BP-RP = -0.2310\,\mathrm{mag}$, consistent with a hot, blue object, and yields an absolute magnitude of $M_G = 7.40\,\pm\,0.21\,\mathrm{mag}$. The nebula spans $9.6'\times 8.3'$, corresponding to physical dimensions of  $2.9^{+0.3}_{-0.3}\,\mathrm{pc} \times 2.5^{+0.3}_{-0.2}\,\mathrm{pc} $ at the \Gaia\ distance. It is important to note that \citet{2026AN....34770078C} independently identified Gaia DR3 1095335102795586944 as the central star of JAM~4/TBG-N1 and spectroscopically confirmed that it is, indeed, a WD. Furthermore, based on the effective temperature and surface gravity they derive, the age of the WD is 100--250 kyr thus providing further support  for the ancient nature of the PN.

\section{Conclusions}

We present narrow-band imaging of three new candidate ancient PNe, which we name JAM~2, JAM~3 and JAM~4.  Morphological analysis indicates that all three have axisymmetric morphologies with strong indications of interaction with the ISM.  Based on the \Gaia\ parallaxes of their candidate central stars (Table \ref{tab:jam_cs}), we estimate their ages to be in the range 50--100 kyr, meaning they are all amongst the oldest PNe known.  The candidate central star of JAM~2 is an N-rich DO WD -- a rare type of WD believed to be the product of binary evolution and the direct progeny of N-rich O(He) stars.  ZTF photometry of the candidate central star of JAM~2 reveals low-amplitude, periodic variability consistent with spots on the rotating stellar surface.  These results demonstrate the potential of modern, advanced amateur astronomy equipment to reveal hitherto hidden ghost PNe.

\begin{acknowledgements}
The authors thank the anonymous referee for their constructive comments.

This work made use of Astropy:\footnote{http://www.astropy.org} a community-developed core Python package and an ecosystem of tools and resources for astronomy \citep{astropy:2013, astropy:2018, astropy:2022}. Based on observations made with the Gran Telescopio Canarias (GTC), installed at the Spanish Observatorio del Roque de los Muchachos of the Instituto de Astrof\'isica de Canarias, on the island of La Palma. This work has made use of data from the European Space Agency (ESA) mission
{\it Gaia} (\url{https://www.cosmos.esa.int/gaia}), processed by the {\it Gaia}
Data Processing and Analysis Consortium (DPAC,
\url{https://www.cosmos.esa.int/web/gaia/dpac/consortium}). Funding for the DPAC
has been provided by national institutions, in particular the institutions
participating in the {\it Gaia} Multilateral Agreement.

DJ acknowledges support from the Agencia Estatal de Investigaci\'on del Ministerio de Ciencia, Innovaci\'on y Universidades (MCIU/AEI) under grant ``Nebulosas planetarias como clave para comprender la evoluci\'on de estrellas binarias'' and the European Regional Development Fund (ERDF) with reference PID2022-136653NA-I00 (DOI:10.13039/501100011033). DJ also acknowledges support from the Agencia Estatal de Investigaci\'on del Ministerio de Ciencia, Innovaci\'on y Universidades (MCIU/AEI) under grant ``Revolucionando el conocimiento de la evoluci\'on de estrellas poco masivas'' and the  European Union NextGenerationEU/PRTR with reference CNS2023-143910 (DOI:10.13039/501100011033). MSG acknowledges support from the I+D+i PID2023-146056NB-C21 (CRISPNESS/MESON), funded by the AEI (10.13039/501100011033) of the Spanish MICIU and the European Regional Development Fund (ERDF) of the EU. NR is supported by the Deutsche Forschungsgemeinschaft (DFG) through grant RE3915/2-1.

\end{acknowledgements}

\bibliographystyle{aa} 
\bibliography{jam}

\end{document}